\documentclass[12pt]{article}

\usepackage[textheight=22.5truecm,textwidth=16.8truecm,top=2.95truecm,left=2.15truecm]{geometry}

\usepackage{amsmath,amsthm,amssymb,ascmac,bm,tikz,braket,mathrsfs,graphicx,color,hyperref,cite,titlesec,authblk,caption,enumitem}

\hypersetup{
  colorlinks,
  bookmarksopen,
  bookmarksnumbered,
  citecolor=black,
  linkcolor=black,
  linktocpage,
  pdfstartview=FitH,
  urlcolor=black
}

\numberwithin{equation}{section}

\def\fnum@figure{\textbf{\figurename\nobreakspace\thefigure}}
\def\fnum@table{\textbf{\tablename\nobreakspace\thetable}}
\long\def\@makecaption#1#2{%
  \vskip\abovecaptionskip
  \sbox\@tempboxa{\small #1. #2}%
  \ifdim \wd\@tempboxa >\hsize
    \small #1. #2\par
  \else
    \global \@minipagefalse
    \hb@xt@\hsize{\hfil\box\@tempboxa\hfil}%
  \fi
  \vskip\belowcaptionskip}
\setlist[enumerate]{nosep,leftmargin=16pt}
\setlist[itemize]{nosep,leftmargin=16pt}

\renewcommand{\d}[0]{\mathrm{d}}

\newcommand{\cmt}[2]{[#1,#2]}
\newcommand{\acmt}[2]{\{#1,#2\}}
\renewcommand{\.}[0]{\hspace{0mm}}

\newcommand{\alpp}[0]{\alpha'}

\allowdisplaybreaks

\title{
  \hfill{\normalsize RIKEN-iTHEMS-Report-26}\\[12pt]
  Holographic Schwinger-Keldysh effective action for heavy quarks in confinement and deconfinement phases
}
\author[1]{Shin Nakamura\thanks{Email: nshin001z@g.chuo-u.ac.jp}}
\author[2]{Daichi Takeda\thanks{Email: daichi.takeda@riken.jp}}
\affil[1]{\textit{Department of Physics, Chuo University, Kasuga, Bunkyo-ku, Tokyo 112-8551, Japan}}
\affil[2]{\textit{iTHEMS, RIKEN, 2-1 Hirosawa, Wako, Saitama 351-0198, Japan}}

\date{}

\begin{document}

\maketitle

\begin{abstract}
The holographic Schwinger-Keldysh (SK) prescription proposed by Skenderis and van Rees (SvR) has the advantage of being applicable whether or not the gravity dual contains a black hole. Taking advantage of this feature, we derive the quadratic effective action for a quark-antiquark pair in the confinement phase within the holographic SK framework of SvR. We also apply the SvR prescription to derive the quadratic effective action for a single heavy quark moving at a constant velocity in a nonequilibrium steady state in the deconfinement phase.
\end{abstract}

\newpage
\tableofcontents
\newpage

\section{Introduction}

The strongly coupled regime of QFT can be explored via the AdS/CFT correspondence\cite{Maldacena:1997re,Gubser:1998bc,Witten:1998qj}, which has deepened our understanding through perturbative calculations in a classical gravitational theory.
In general it is difficult to study strongly coupled regions of QFT, such as the confining phase of QCD.
However, by following the prescriptions of AdS/CFT one can compute QFT correlators by evaluating the on‑shell action in the bulk to the necessary order.

Many techniques have been proposed for analyzing finite temperature QFT using AdS/CFT.
A prescription for obtaining retarded Green’s functions in a Lorentzian signature bulk was first proposed in \cite{Son:2002sd}.
To obtain multipoint correlation functions one must introduce two real time contours and work within the Schwinger-Keldysh (SK) formalism.
For this purpose, \cite{Herzog:2002pc,Son:2009vu} maximally extended a static black hole spacetime and used the two AdS boundaries.
The bulk fields are analytically continued across the horizon following Unruh’s prescription\cite{Unruh:1976}.
More recently, \cite{Glorioso:2018mmw} proposed a new analytic continuation across the horizon, which has attracted considerable attention because of its convenience.

Although there are many proposals as described above, a prescription for general real and imaginary time evolution (including the SK contour) was first systematically developed by Skenderis and van Rees (SvR)\cite{Skenderis:2008dh,Skenderis:2008dg,vanRees:2009rw}.\footnote{Recently, \cite{Ammon:2025vod} proposed a treatment for more general situations.}
Their method has two main advantages:
\begin{enumerate}[nosep]
  \item It is derived from the “equivalence of path integrals” between the boundary and the bulk,
  \item It is applicable irrespective of the presence or absence of a horizon.
\end{enumerate}
Point a) is important because it provides, quite literally, a proof based on the most fundamental assumption of AdS/CFT.
In particular, the prescription of \cite{Son:2002sd} is justified within the SvR prescription \cite{vanRees:2009rw}.
Point b) matters because if one is to consider the low‑temperature phase of the Hawking-Page transition \cite{HawkingPageTransition,Witten:1998zw,Horowitz:1998ha,Surya:2001vj}, the prescription must remain valid even in the absence of a horizon.\footnote{Applications of the SvR prescription are numerous; those that treat the low‑temperature phase in the in‑in formalism include \cite{Skenderis:2008dg,Botta-Cantcheff:2018brv}.}~\footnote{\label{foot:Son-Teaney}Even for the low‑temperature phase, the method of \cite{Son:2002sd} can be used by avoiding poles on the real axis in Fourier space along a retarded contour. However, one should keep in mind that the method of \cite{Son:2002sd} cannot be applied to multipoint correlation functions.}
The fact that one can dispense with a horizon implies that there is no need to introduce certain analytic continuation across the horizon.

The main purpose of this paper is to exploit the advantages of the SvR prescription to study the fluctuation and dissipation Green functions of a quark-antiquark pair in the confinement phase.
At zeroth order in the perturbation the configuration is a static test quark-antiquark pair, whose bulk dual is the U-shaped static Nambu-Goto string (Fig.\ \ref{fig: string in soliton})\cite{Maldacena:1998im,Brandhuber:1998er}.
We then consider arbitrary but small deformation to the worldline of one of the quarks away from rest.
The response Green function to this deformation is obtained by constructing the SK contour in the bulk following the SvR prescription and evaluating the on‑shell action to quadratic order.
In the confinement phase, the Green function has poles in frequency space, and these poles carry information about the excitation modes of the flux tube.

We also use the SvR prescription to investigate the Green function for a single test quark moving at constant velocity in the deconfinement phase.
The zeroth‑order solution in the perturbation is the one constructed in \cite{Herzog:2006gh,Gubser:2006bz}; the trailing open string that has one endpoint attached to the boundary and moves at constant velocity while the other endpoint dumps energy into the horizon (see also \cite{Casalderrey-Solana:2006fio,Gubser:2006nz,Casalderrey-Solana:2007ahi}).

Because there exist many previous studies in the deconfinement phase using prescriptions other than the SvR prescription, we briefly mention some representative works.
Earlier studies of the stochastic dynamics of a heavy quark include \cite{Son:2009vu,Giecold:2009cg,Casalderrey-Solana:2009ifi}.
More recently, \cite{Bu:2021jlp} employed the prescription of \cite{Glorioso:2018mmw} to derive the Schwinger-Keldysh effective action for a trailing string on a black-brane background up to third order in perturbations.
Another approach to analyzing the Brownian motion is to quantize the string on black hole backgrounds \cite{deBoer:2008gu,Lawrence:1993sg}.

This paper is organized as follows:
\begin{itemize}[nosep]
  \item Sec.\ \ref{sec: SK review}: We review thermal field theory and the SvR prescription.
  \item Sec.\ \ref{sec: low temp}: We determine the bulk worldsheet corresponding to a static quark-antiquark pair on the AdS$_5$ soliton spacetime and use it to construct the SvR spacetime (Fig.~\ref{fig: SvR spacetime} left). We perturb one endpoint of the string and evaluate the on‑shell action to second order.
  \item Sec.\ \ref{sec: high temp}: We determine the bulk worldsheet corresponding to a single quark trailing at constant velocity on the AdS$_5$ black‑brane spacetime and use it to construct the SvR spacetime (Fig.~\ref{fig: SvR spacetime} right). We perturb the endpoint of the string on the boundary and evaluate the on‑shell action to second order.
  \item Sec.\ \ref{sec: discussion}: We devote this section to summary and discussion.
  \item App.\ \ref{app: NG expansion}: We derive the perturbative expansion formula for the Nambu-Goto action used in Sec.\ \ \ref{sec: low temp}.
  \item App.\ \ref{app: junction}: In Sec.\ s\ \ref{sec: high temp} and \ref{sec: low temp} we glue Lorentzian and Euclidean spacetimes according to the SvR prescription. This gluing contains subtle technical points not addressed in previous work; we carefully treat these details in this appendix.
  \item App.\ \ref{app: scalar}: We calculate the SK effective action of a bulk probe free scalar field on a black hole background.
\end{itemize}

In Sec.\ \ref{sec: low temp} we obtain Green functions, but they are formally presented by using the normal modes of the bulk equations of motion and the coefficients read off from the asymptotic expansion of solutions.
Determining these coefficients requires a detailed numerical analysis of the equations of motion, which we leave for future work (see also Sec.\ \ \ref{sec: discussion}).
The main task of this paper is to implement the SK formalism for quark (pairs) holographically using the SvR prescription.

\section{Review: Schwinger-Keldysh formalism and holography}\label{sec: SK review}

In this section, we review the thermal QFT based on the Schwinger-Keldysh (SK) formalism\footnote{For a more detailed review, we recommend Ref.~\cite{Liu:2018kfw}.}, and its holographic realization developed by Skenderis and van Rees (SvR).  

\subsection{Thermal fields in the boundary theory}\label{subsec: thermal field}
\begin{figure}[t]
	\centering
	\includegraphics[height = 4.5cm]{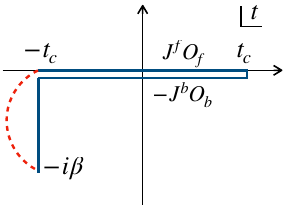}
	\caption{The Schwinger-Keldysh contour $C_{\mathrm{SK}}$.  
	The blue line represents the contour, whose initial and final points are periodically identified due to the cyclic property of the trace.}
	\label{fig: SK contour}
\end{figure}

Here, we explain the SK contour in QFT and see how various Green functions can be computed.
For simplicity, we focus on a real scalar operator $O(x)$ in the boundary theory, and later we discuss the coupling between a probe particle and a gauge field, which will be the main subject of this paper.

The generating functional for thermal correlation functions is given by  
\begin{align}\label{eq: SK in operator}
  Z[J] :=& \mathrm{Tr}\left[e^{-\beta H}\left\{\bar{ \mathrm{T}}e^{i\int \d^d x\, (H - J^bO)} \right\} \left\{\mathrm{T}e^{-i\int\d^d x\, (H - J^fO)} \right\} \right],
\end{align}
where $\beta$ is the inverse temperature, $H$ is the Hamiltonian, $\mathrm{T}$ ($\bar{\mathrm{T}}$) denotes (anti-)time ordering, and the sources $J^f$ and $J^b$ have support on $t \in (-t_c, t_c)$, with $t_c > 0$ being a time cutoff.  
Rewriting \eqref{eq: SK in operator} in the path-integral representation,\footnote{To express it as a path integral, we assume that $O(x)$ does not depend on the conjugate momenta of the fundamental fields.} we obtain  
\begin{align}\label{eq: SK in path int}
  Z[J] = & \oint \mathcal{D}\phi\, e^{i I[\phi; C_{\mathrm{SK}}]+i \int \d^d x\, (J^fO_f - J^bO_b)}.
\end{align}
Here, $\phi$ collectively denotes the fundamental fields of the theory, $C_\mathrm{SK}$ refers to the contour shown in Fig.~\ref{fig: SK contour}, and $O_f = O(t,\bm x)$, $O_b = O(t-i0,\bm x)$.  
This path-integral expression will play a crucial role when we apply the AdS/CFT correspondence in Sec.~\ref{subsec: SvR}.

The Green functions are defined as follows (with $\rho_\beta := e^{-\beta H}/\mathrm{Tr}\, e^{-\beta H}$):  
\begin{align}\label{eq: def Green functions}
  G_\mathrm{R}(x_1,x_2) &:= -i\Theta(t_1-t_2)\mathrm{Tr}\left(\rho_\beta \cmt{O(x_1)}{O(x_2)} \right),\nonumber\\
  G_\mathrm{A}(x_1,x_2) &:= -i\Theta(t_2-t_1)\mathrm{Tr}\left(\rho_\beta \cmt{O(x_2)}{O(x_1)} \right),\nonumber\\
  G_\mathrm{S}(x_1,x_2) &:= \frac{1}{2}\mathrm{Tr}\left(\rho_\beta \acmt{O(x_1)}{O(x_2)} \right).
\end{align}
Since $\rho_\beta$ commutes with the momentum operator $P_\mu = (-H,\bm P)$, all of these Green’s functions depend only on $x_1 - x_2$.  
Accordingly, we define their Fourier transforms as  
\begin{align}
	G_l(k) = \int \d^d x\, e^{-ik\cdot x} G_l(x),\qquad (l = \textrm{R},\, \textrm{A},\, \textrm{S}).
\end{align}

In the so-called \textit{retarded-advanced (ra)} basis, we can easily extract the retarded, advanced, and symmetric Green’s functions.
In \eqref{eq: SK in operator}, by redefining the sources as  
\begin{align}\label{eq: ra sources}
  J^r = \frac{1}{2}(J^f + J^b),\qquad
  J^a = J^f - J^b,
\end{align}
we find that  
\begin{align}\label{eq: Green functions by Z}
  G_\mathrm{R}(x_1,x_2) &= \frac{i}{Z_0[0]}\left.\frac{\delta^2}{\delta J^a(x_1)\delta J^r(x_2)}Z[J]\right|_{J\to 0},
  \nonumber\\
  G_\mathrm{S}(x_1,x_2) &= -\frac{1}{Z_0[0]}\left.\frac{\delta^2}{\delta J^a(x_1)\delta J^a(x_2)}Z[J]\right|_{J\to 0}.
\end{align}
To verify these expressions, one can perform the functional differentiations of $Z[J]$ using \eqref{eq: SK in operator} and \eqref{eq: ra sources}, and then compare the results with the definitions given in \eqref{eq: def Green functions}.

\subsubsection*{Probe quarks}

In Sec.~\ref{sec: low temp} and thereafter, we investigate the thermal properties of the motion of probe quarks.
In this setup, the particle’s position is treated as an external variable, which couples to the gauge field of the QFT.  
Such a situation can be analyzed by means of the Wilson loop:
\begin{align}\label{eq: Wilson loop}
  Z[x(\cdot)] = \oint \mathcal{D}A\, e^{i I[A;C_{\mathrm{SK}}] + \int \d^dx\, j^\mu A_\mu},
  \qquad
  j^\mu(x) := \oint_W \d s\, \dot x^\mu(s)\, \delta(x - x(s)).
\end{align}
Here, $W$ denotes the Wilson loop contour, and $x^\mu(s)$ is its parametrization.  
For simplicity, we give the expression for a $\mathrm{U}(1)$ gauge group, but the generalization to a non-Abelian gauge field is straightforward.

In this case, varying the source corresponds to a deformation of the path, $x^\mu(s) \to x^\mu(s) + \delta x^\mu(s)$.  
For instance, by performing a single variation, one can measure the force exerted on the probe particle by the gauge field:
\begin{align}\label{eq: Field strength}
  Z[x(\cdot) + \delta x(\cdot)] - Z[x(\cdot)]
  = i \oint \d s\, \delta x^\mu(s) \oint \mathcal{D}A\, e^{i I[A;C_{\mathrm{SK}}] + \int \d^dx\, J^\mu A_\mu}\, F_{\mu\nu}(x(s))\, \dot x^\nu(s).
\end{align}
Here, $F_{\mu\nu}$ is the field strength, and the combination $F_{\mu\nu}\dot x^\nu$ has the same form as the one that appears in the equation of motion for a relativistic charged particle.

\subsection{The prescription by Skenderis and van Rees}\label{subsec: SvR}

\begin{figure}[t]
	\centering
  \begin{minipage}{0.45\columnwidth}
    \centering
    \includegraphics[height = 7cm]{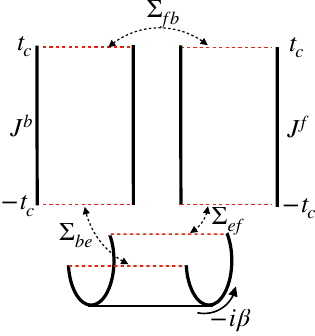}
  \end{minipage}
  \begin{minipage}{0.45\columnwidth}
    \centering
    \includegraphics[height = 7cm]{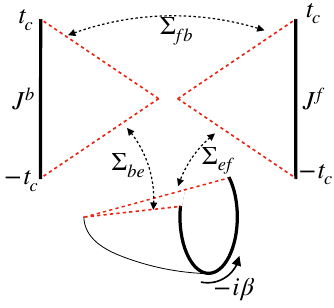}
  \end{minipage}
  \caption{The spacetime $M_{\mathrm{SvR}}$ dual to the SK contour (Fig.~\ref{fig: SK contour}).  
  $M_{f,b}$ are Lorentzian, while $M_e$ is Euclidean, and $\Sigma_{kl}$ denotes the junction surface connecting $M_k$ and $M_l$.  
  The time path along the boundary (thick lines) reproduces exactly the SK contour.  
  The left panel represents the low temperature phase, where there is no horizon, and the diagram is drawn like the Penrose diagram of global AdS.  
  The right panel corresponds to the high temperature phase.  
  Although $M_f$ and $M_b$ appear as if they were parts of a maximally extended black hole spacetime, they are in fact two independent single-sided black hole spacetime.}
   \label{fig: SvR spacetime}
\end{figure}

We now review the holographic SK formalism based on the method of Skenderis and van Rees (SvR)\cite{Skenderis:2008dh,Skenderis:2008dg,vanRees:2009rw}.  
To this end, we assume that the action $I$ in \eqref{eq: SK in path int} is that of a holographic CFT, and that the operator $O$ is a scalar primary operator of conformal dimension $\Delta$.

Let $S$ denote the bulk action dual to this holographic CFT.  
Then we have
\begin{align}\label{eq: bulk Z}
  Z[J] &= \oint \mathcal{D}\Phi\, e^{i S[\Phi; M_{\mathrm{SvR}}]}
  \,\simeq\, e^{iS_{\mathrm{cl}}[\Phi; M_{\mathrm{SvR}}]},\\
  \mbox{with}\quad
  &\Phi_f \sim r^{\Delta-d} J^f,\quad
  \Phi_b \sim r^{\Delta-d} J^b,\quad
  \Phi_e \sim 0\quad (\mbox{as }r\to\infty),
  \label{eq: bulk BC}
\end{align}
where $M_{\mathrm{SvR}}$ is a spacetime whose AdS boundary traces the SK contour shown in Fig.~\ref{fig: SK contour}, as illustrated in Fig.~\ref{fig: SvR spacetime}.  
Here, $f$ and $b$ label Lorentzian segments, and $e$ labels the Euclidean segment.  
The field $\Phi$ is a real scalar field with mass squared $m^2 = \Delta(\Delta - d)/L^2$, and \eqref{eq: bulk BC} gives the boundary conditions on each segment of $M_{\mathrm{SvR}}$.  
The subscripts $f,b,e$ on $\Phi$ correspond to $M_f$, $M_b$, and $M_e$ in Fig.~\ref{fig: SvR spacetime}.  
The fields are glued across the junction surfaces $\Sigma_{kl}$ connecting $M_k$ and $M_l$.  
Below we treat the scalar field as a probe.

To construct classical solutions on the SvR spacetime, one must also impose junction conditions at the interfaces $\Sigma_{kl}$.  
These junction conditions arise from taking the saddle-point approximation of the path integral \eqref{eq: bulk Z}.  
The result can be stated as follows, denoting by $\Pi$ the canonical momentum conjugate to $\Phi$:
\begin{align}\label{eq: junction conditions}
  \Sigma_{fb}: \Phi_f = \Phi_b~\&~\Pi_f = \Pi_b,\quad
  \Sigma_{be}: \Phi_b = \Phi_e~\&~\Pi_b = i\Pi_e,\quad
  \Sigma_{ef}: \Phi_e = \Phi_f~\&~\Pi_f = i\Pi_e.
\end{align}
We will briefly explain their origin at the end of this subsection.

Of course, similar junction conditions must also be imposed for the gravitational field.  
In this case, in \eqref{eq: junction conditions} one should replace $\Phi$ by the induced metric on $\Sigma_{kl}$ and $\Pi$ by its extrinsic curvature tensor.  
However, since we treat the scalar field as a probe, we adopt a stationary vacuum solution for the metric throughout $M_{\mathrm{SvR}}$.  
If we choose the junction surfaces $\Sigma_{kl}$ to be constant-time slices determined by the timelike Killing vector, then the gravitational junction conditions are automatically satisfied.

To understand the origin of \eqref{eq: junction conditions}, let us consider a much simpler system: a one-dimensional quantum system whose SK path integral is approximated by a saddle point.  
The action is
\begin{align}\label{eq: QM action}
  i I_{\mathrm{QM}} = i\int_{-t_c}^{t_c}\d t\,\left(\frac{(\partial_t q_f)^2}{2} - V(q_f)\right)
  - \int_0^{\beta}\d \tau\,\left(\frac{(\partial_\tau q_e)^2}{2} + V(q_e)\right) + \cdots,
\end{align}
where we have explicitly written only a part of the SK contour.  
The path integral and its saddle-point approximation correspond to \eqref{eq: bulk Z} with $S \to I_{\mathrm{QM}}$.  
We focus on the junction at $t = -t_c$ and $\tau = \beta$.  
In taking the saddle-point approximation, one must consider that  
\begin{enumerate}[nosep]
  \item $q_f(-t_c)$ and $q_e(\beta)$ take the same value, denoted $q_{ef}$, and  
  \item $q_{ef}$ itself is not fixed.  
\end{enumerate}
Point a) corresponds to $\Phi_e = \Phi_f$ in \eqref{eq: junction conditions}, so let us see how the momentum condition appears.
With point a) and b) in mind, varying \eqref{eq: QM action} yields
\begin{align}
  \delta(i I_{\mathrm{QM}}) &=  -i\int_{-t_c}^{t_c}\d t\,\left(\partial_t^2 q_f + V'(q_f)\right)\delta q_f
  + \int_0^{\beta}\d \tau\,\left(\partial_\tau^2 q_e - V'(q_e)\right)\delta q_e \nonumber\\
  &\hspace{24pt}
  + \left[-i \partial_t q_f(-t_c) - \partial_\tau q_e(\beta)\right] \delta q_{ef} + \cdots.
\end{align}
The first line gives the equations of motion for the saddle-point solution, while the second line (the surface term) determines the junction condition.  
We cannot impose the Dirichlet condition $\delta q_{ef} = 0$ due to point b).
Therefore, the saddle-point solution must satisfy at the junction surface,
\begin{align}
   \partial_t q_f(-t_c) = i\,\partial_\tau q_e(0).
\end{align}
This corresponds to $\Pi_f = i\Pi_e$ in \eqref{eq: junction conditions}.  
For a detailed derivation of \eqref{eq: junction conditions} for scalar and gravitational fields, see \cite{Skenderis:2008dg}.

In appendix \ref{app: scalar}, we carry out the computation and obtain the effective action in the free scalar theory.

\subsubsection*{Probe quarks}

For the gauge field coupled to probe quarks as in \eqref{eq: Wilson loop}, the dual object is the Nambu-Goto string\cite{Maldacena:1998im},
\begin{align}
  Z[x(\cdot)] \simeq e^{iS_{\mathrm{NG}}[X; \partial X = W]},
\end{align}
with the action in Lorentzian signature given by
\begin{align}\label{eq: NG action}
  S_{\mathrm{NG}} = -\int \d^2\sigma\, \sqrt{-\gamma},\qquad
  \gamma_{ab} = G_{\mu\nu}(X)\, \partial_a X^\mu\, \partial_b X^\nu.
\end{align}  
Here, we have set $2\pi\alpp = 1$, $G_{\mu\nu}$ is the bulk metric, and
\begin{align}\label{eq: world sheet}
  X^\mu = (t, r, X^1(t,r), \ldots, X^{d-1}(t,r))
\end{align}
represents the worldsheet.  
The worldsheet coordinates are gauge-fixed to the background coordinates $(t,r)$, i.e., $\sigma^a = (t,r)$.  
The boundary condition $\partial X = W$ means that the boundary of the worldsheet lies on the Wilson loop $W$.

In this paper, we are interested in trajectories $W$ that are timelike in their Lorentzian segments.  
To close the loop, one must go once around the SK contour $C_{\mathrm{SK}}$.  
In the low-temperature phase, we consider a quark-antiquark pair, so there are two such loops;  
in the high-temperature phase, we consider a single quark, so there is one.

When constructing the string solution on $M_{\mathrm{SvR}}$, one also needs to impose junction conditions.  
To do this, one simply replaces $\Phi$ and $\Pi$ in \eqref{eq: junction conditions} by $X^1,\ldots,X^{d-1}$ and their conjugate momenta.

\section{Confinement phase: a quark-antiquark pair}\label{sec: low temp}
\begin{figure}
	\centering
    \includegraphics[height = 4cm]{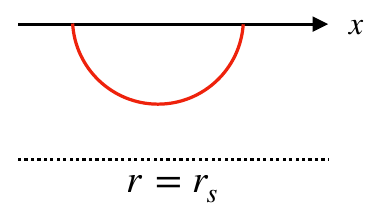}
  \caption{
  Static string in the AdS$_5$ soliton phase.
  The red curve depicts the string profile.
  As inferred from \eqref{eq: Lorentz soliton}, the neighborhood of $r=r_s$ is a Euclidean plane.}
   \label{fig: string in soliton}
\end{figure}

In this section, we consider the static configuration of a Nambu-Goto string in the confinement phase (the AdS$_5$ soliton) and evaluate the on-shell action to quadratic order in perturbations around it.
From the viewpoint of a constant-time slice in the bulk, the static configuration is a straight string stretched along the time direction with both endpoints attached to the boundary (see Fig.~\ref{fig: string in soliton}).
Starting from that configuration, we slightly displace the coordinate of one endpoint, and we solve for the perturbed solution on the SvR spacetime (Fig.~\ref{fig: SvR spacetime}, left).

\subsection{Static solution}\label{subsec: both endpoints}

Following \cite{Maldacena:1998im, Brandhuber:1998er}, we determine the string dual to a static quark-antiquark pair (Lorentz signature).
The AdS$_5$ soliton background in the bulk is
\begin{align}\label{eq: Lorentz soliton}
  \d s^2 &=  -\frac{r^2}{L^2}\left(-\d t^2 + (\d x^1)^2+ (\d x^2)^2+ h(r)(\d x^3)^2 \right) + \frac{L^2}{r^2h(r)}\d r^2,
  \qquad
  \left(h(r) := 1 - \frac{r_s^4}{r^4} \right)
\end{align}
where $L$ is the AdS radius and $r_s$ is a positive constant.
For \eqref{eq: world sheet}, we impose the ansatz
\begin{align}\label{eq: soliton ansatz}
  X^1 = a-\xi(r),\qquad X^{2} = X^3= 0,\qquad
  \xi(\infty) = 0,\qquad
  \xi'(r_*) = -\infty,
\end{align}
which describes only the right half of the string; the full solution is obtained by reflection.
The constants $a$ and $r_*\, (>r_s)$ are related by $\xi(r_*)=a$, and below we regard $a$ as determined by $r_*$.

Under \eqref{eq: soliton ansatz}, the Lagrangian is given by
\begin{align}
  \mathcal{L}_0 = -\sqrt{\frac{1}{h(r)}+\frac{r^4 \xi'(r)^2}{L^4}},
\end{align}
so $\xi$ is a cyclic coordinate.
Integrating the EOM once and solving for $\xi'(r)$, we obtain
\begin{align}\label{eq: xi prime in soliton}
  \xi'(r) = - \frac{r_*^2L^2}{\sqrt{(r^4-r_*^4)(r^4-r_s^4)}},
\end{align}
where the integration constant is fixed by the last condition in \eqref{eq: soliton ansatz}.
A further integration yields an explicit expression for $\xi(r)$, but we do not need it.

For this solution, the induced worldsheet metric is
\begin{align}\label{eq: pre soliton world sheet}
  \d s^2_\mathrm{WS} = -\frac{r^2}{L^2}\d t^2 + \frac{L^2r^6}{(r^4-r_*^4)(r^4-r_s^4)}\d r^2.
\end{align}
Although this metric is defined for $r\in (r_*, \infty)$, the new coordinate
\begin{align}\label{eq: rho coordinate}
  \rho := \int_{r_*}^{r}\d\tilde r\;\frac{L^2\tilde r^2}{\sqrt{(\tilde r^4-r_*^4)(\tilde r^4-r_s^4)}}
  \sim 
  \begin{cases}
  	\rho_\infty - L^2/r + \mathcal{O}(r^{-5}) & (r\to \infty)\\
  	(\mathrm{const})\cdot \sqrt{r-r_*} & (r\to r_* + 0)
  \end{cases},
\end{align}
resolves the coordinate singularity:
\begin{align}\label{eq: soliton world sheet}
  \d s^2_\mathrm{WS}  = \frac{r(|\rho|)^2}{L^2}(-\d t^2 + \d \rho^2) =: \gamma_{ab}\d \sigma^a \d \sigma^b.
\end{align}
Here $\rho_\infty := \rho(r\to \infty)$ is a finite constant, and $r(\rho)$ is obtained by inverting \eqref{eq: rho coordinate}.
Below we use $(t,\rho)$ as worldsheet coordinates, extend to $\rho<0$, and regard $\rho>0$ as the right half of the string and $\rho<0$ as the left half (this is why we wrote $|\rho|$ in \eqref{eq: soliton world sheet}).

\subsection{Quadratic effective action for perturbations}\label{subsec: quadratic action low temp}
From the boundary perspective, the static string above represents a test quark-antiquark pair coupled to the CFT and forming thermal equilibrium with it.
In this subsection, we perturb the right quark to drive the system out of equilibrium and compute the fluctuation and dissipation Green’s functions.
To that end we must set up the SK contour in the bulk, namely the SvR spacetime (Fig.~\ref{fig: SvR spacetime}, left).

The junction conditions in \eqref{eq: junction conditions} are imposed order by order in perturbations.
In the probe limit, the background geometry is the AdS$_5$ soliton throughout $M_{f,b,e}$ (Euclidean signature on $M_e$).
The zeroth order in perturbations is the static solution of the previous subsection, and, as for the background geometry, the junction conditions are satisfied simply by analytically continuing time in \eqref{eq: soliton world sheet}.

Now, we slightly shift the endpoint (the position of the right quark) in $x^1$-direction as
\begin{align}\label{eq: soliton perturbed BC}
  X^1_f(t,\rho_\infty) &= a  + J^f,\quad
  X^1_b(t,\rho_\infty) = a  + J^b,\quad
  X^1_e(t,\rho_\infty) = a,\\
  X^1_f(t,-\rho_\infty) &=  
  X^1_b(t,-\rho_\infty) = 
  X^1_e(t, -\rho_\infty) = -a.
\end{align}
We solve the EOM upto $\mathcal{O}(J^1)$, imposing \eqref{eq: junction conditions}.

For the perturbative analysis, it is convenient to choose the perturbation in the direction normal to the worldsheet \eqref{eq: soliton world sheet}:\footnote{
We place the indices $f,b,e$ upstairs on $\Phi$ merely because we will later write the Fourier modes as $\Phi^{f,b,e}_{\omega}$.
}
\begin{align}\label{eq: normal perturbation}
  \delta X_{f,b,e}^\mu = \Phi^{f,b,e}(t,\rho) n^\mu,\qquad
  n^\mu := \frac{L}{r^3}\left(\sqrt{r^4-r_*^4}\delta_1^\mu - \frac{r_*^2}{L^2}\sqrt{r^4 - r_s^4}\delta_r^\mu \right).
\end{align}
Here $n^\mu$ is the unit normal restricted to the plane $x^{2}=x^3=0$, and for $\rho<0$ we take $n^x<0$.
In \eqref{eq: soliton perturbed BC}, for example $X^1_{f}(t,\rho_\infty) = a + J^f$ is replaced by
\begin{align}\label{eq: rho BC}
  &\Phi^f(t,\rho_\infty - \delta) = \frac{R}{L} J^f(t),\\
  &\delta := \int_R^{\infty}\d r\; \frac{L^2 r^2}{\sqrt{(r^4-r_*^4)(r^4-r_s^4)}}\overset{R\to \infty}{\sim } \frac{L^2}{R} + \mathcal{O}(R^{-5}).
\end{align}
See App.~\ref{app: NG expansion} for the detail.
Here $r=R$ denotes the location of the cutoff surface near the boundary, and we take $R\to\infty$ after evaluating the action.
From \eqref{eq: rho coordinate}, this cutoff surface is located at $\rho=\rho_\infty-\delta$ in the $\rho$ coordinate.

As derived in App.~\ref{app: NG expansion}, the perturbative expansion of the action on $M_f$ is given by
\begin{align}\label{eq: soliton action expansion}
  S_\mathrm{NG}[X_f] &= (\mathrm{const}) + \int_{-t_c}^{t_c}\d t \left(-\frac{r_*^2}{L^2}J^f(t) + \frac{R^3}{2L^4}J^f(t)^2  \right)\nonumber\\
  &\hspace{24pt}
  -\frac{1}{2}\int_{-t_c}^{t_c}\d t \int_{-\rho_\infty + \delta}^{\rho_\infty - \delta}\d \rho\sqrt{-\gamma}\left(\gamma^{ab}\partial_a \Phi \partial_b \Phi + m^2(r)\Phi^2 \right).
\end{align}
Here, we have defined
\begin{align}
  m^2(r) &:= -R_{\mu \nu }\.^\rho\._\sigma e^\mu_a n^\nu e_\rho^a n^\sigma - K^{ab}K_{ab}\label{eq: mass of r}
  \\
  &= \frac{2}{L^2}\frac{r^{12}-5r_*^4 r^8 + 12r_*^8 r^4 - 8r_*^{12} - r_s^4(r^8 + r_*^4 r^4 - 4 r_*^8) - 2 r_*^4 r_s^8}{(r^4 - r_s^4)r^8}\\
   &=
  \begin{cases}
  	2/L^2 & (\rho = \pm \rho_\infty)\\
  	4r_s^4/(L^2r_*^4) & (\rho = 0)
  \end{cases},
\end{align}
where $R_{\mu\nu }\.^\rho\._\sigma$ is the Riemann tensor of the AdS$_5$ soliton, $K_{ab}$ is the extrinsic curvature of the static string along $n^\mu$, and $e_a^\mu$ is defined from the zeroth-order solution \eqref{eq: soliton ansatz} as $e_a^\mu : = \partial X^\mu/\partial \sigma^a$ and $e_\mu^a := g_{\mu\nu }\gamma^{ab}e_b^\nu$.
Replacing $J^f\to J^b$ gives $S_{\mathrm{NG}}[X_b]$, and setting $J^f\to 0$ with $t\to -i\tau$ ($\tau\in[0,\beta]$) gives $S_\mathrm{NG}[X_e]$.

\subsubsection{EOM and general solution in Lorentz signature}\label{subsubsec: Lorentz EOM}
We first analyze the EOM for $\Phi^f$, which will then apply to the other segments of the SvR spacetime.

\begin{figure}[t]
	\centering
	\includegraphics[height = 7cm]{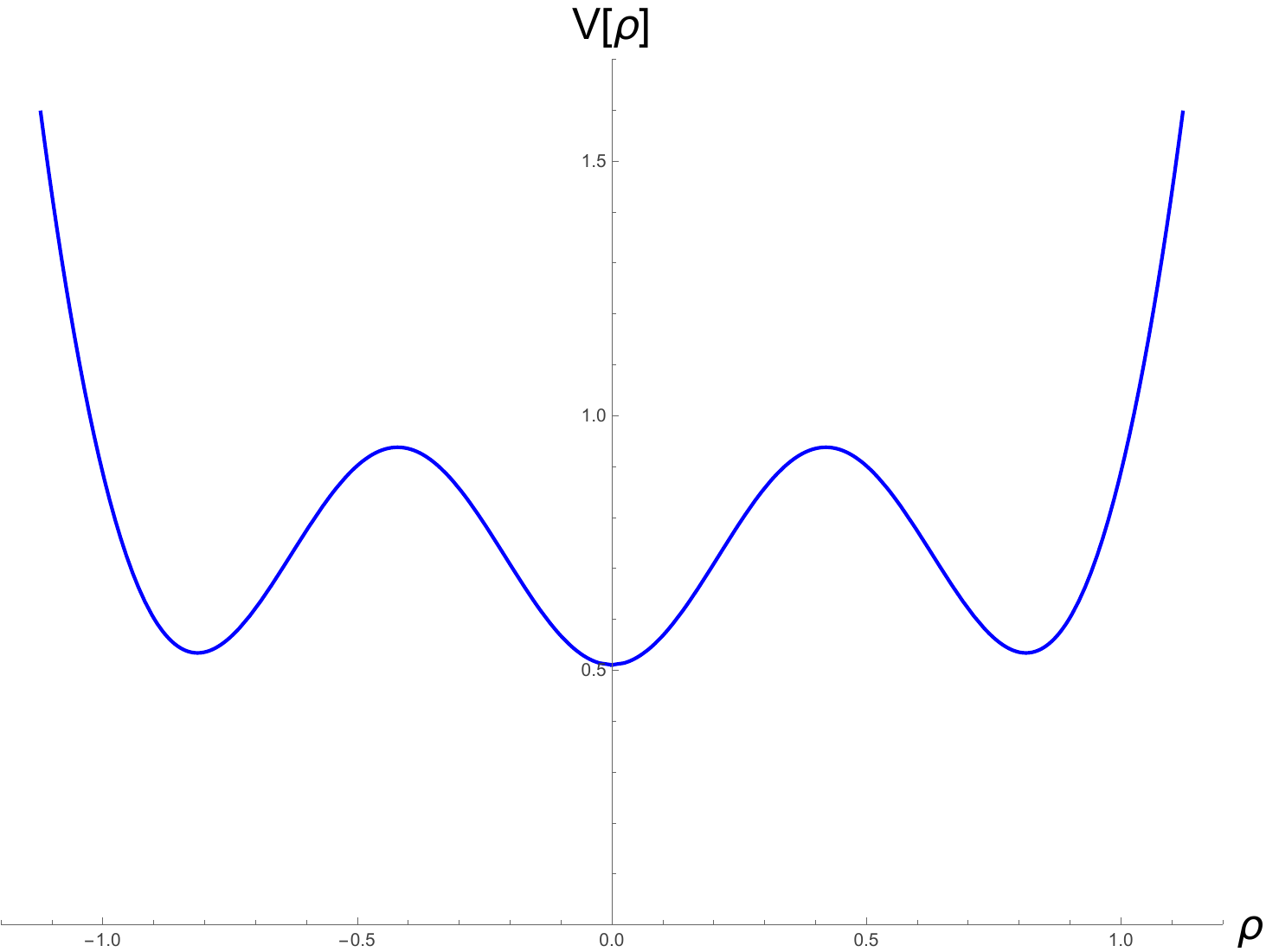}
	\caption{Shape of the potential $V(\rho)$ in \eqref{eq: Schrodinger}, with $L = 1$, $r_s = 0.7$, and $r_* = 0.5$.}
	\label{fig: potential}
\end{figure}

With the Fourier transform
\begin{align}\label{eq: Fourier tr}
  \Phi^f(t,r) = \int \frac{\d\omega}{2\pi}\, e^{-i \omega t} \Phi^f_{\omega}(r),
\end{align}
the EOM becomes Schr\"odinger type:\footnote{
Although not needed below, the potential $V(\rho)$ is smooth: the continuity at $\rho=0$ is obvious, and $V'(\rho)\sim \rho\, \d V/\d r$ is also continuous because $\d V/\d r$ is finite at $r=r_*$ (using the expansion of \eqref{eq: rho coordinate}).
}
\begin{align}\label{eq: Schrodinger}
  \left(-\frac{\d^2}{\d \rho^2} + V(\rho) \right) \Phi^f_{\omega} = \omega^2 \Phi^f_{\omega},\qquad
  V(\rho) := \frac{r(|\rho|)^2}{L^2}m^2(r(|\rho|)).
\end{align}
Fig.~\ref{fig: potential} shows the shape of the potential $V(\rho)$.

From \eqref{eq: Schrodinger} near $\rho\to \pm \rho_\infty$, we find
\begin{align}\label{eq: soliton asymptotic}
  \Phi^f_\omega \sim |\rho \mp \rho_\infty|^{-1},~|\rho \mp \rho_\infty|^{2}
  \qquad
  \left(\rho\to \pm \rho_\infty \right),
\end{align}
namely a decaying solution and a divergent one.

Solutions decaying at both ends are bound states in the sense of quantum mechanics.
Because the potential is positive and diverges at both ends, there are countably infinite number of bound modes $\{\pm \omega_n\,|\,n=0,1,2,\dots\}$, where we choose $\omega_n>0$.
According to quantum mechanics, there is no degeneracy in one dimension.

In view of \eqref{eq: soliton perturbed BC} (see also \eqref{eq: rho BC}), we also need a solution that diverges at the right end and decays at the left.
Let $\hat\phi_\omega$ satisfy
\begin{align}
  \lim_{\rho\to -\rho_\infty} \hat \phi_\omega (\rho) (\rho + \rho_\infty)^{-2} = 1,\qquad
  \lim_{\rho\to -\rho_\infty} \hat \phi_\omega' (\rho) (\rho + \rho_\infty)^{-1} = 2,
\end{align}
so $\hat \phi_\omega (\rho) \sim (\rho + \rho_\infty)^2$.
In general, near the right end we can expand it as
\begin{align}
  \hat \phi_\omega(\rho) \overset{\rho \to  \rho_\infty}{\sim}  A(\omega)(\rho_\infty - \rho)^{-1} \left(1 + \frac{\omega^2}{2}(\rho_\infty-\rho )^2 + \cdots \right) + B(\omega)(\rho_\infty - \rho)^2  \left(1+ \cdots \right).
\end{align}
As stated, precisely at $\omega=\pm \omega_n$ the solution becomes a bound state, so $A(\omega)$ has zeros at $\omega=\pm \omega_n$.
Conversely, there are no other bound modes, hence the zeros of $A(\omega)$ lie on the real axis.

Both the divergent mode and the bound modes can be handled uniformly by defining
\begin{align}
  \phi_\omega(\rho) &:= \frac{L}{A(\omega)}\hat \phi_\omega(\rho),\nonumber\\
	&\overset{\rho \to  \rho_\infty}{\sim}  L(\rho_\infty - \rho)^{-1} \left(1 + \frac{\omega^2}{2}(\rho_\infty-\rho )^2 + \cdots \right) + \frac{LB(\omega)}{A(\omega)}(\rho_\infty - \rho)^2  \left(1+ \cdots \right),\label{eq: phi expansion}
\end{align}
which has simple poles at the bound modes on the $\omega$-plane (nondegenerate), and is otherwise regular.
Therefore, contour integration around a pole extracts the bound state:\footnote{
For the second equality, note that $A(\omega)$ is even by definition and behaves as $\omega^2-\omega_n^2$ near each zero.
}
\begin{align}\label{eq: def of N}
  N_n(\rho) := -\oint_{\omega_n} \frac{\d\omega}{2\pi }\phi_\omega(\rho) = \oint_{-\omega_n}\frac{\d \omega}{2\pi}\phi_\omega(\rho)
  \overset{\rho \to  \rho_\infty}{\sim}
  -i L \frac{B(\omega_n)}{A'(\omega_n)}(\rho_\infty-\rho)^2(1+\cdots).
\end{align}

So far we have classified solutions by their behavior at $\rho\to\pm \rho_\infty$.
To regularize at a finite cutoff and take the limit $R\to\infty$ (i.e., $\delta\to 0$) carefully, we now reclassify solutions by boundary conditions at $\rho=\pm(\rho_\infty-\delta)$ and relate them to the $R\to\infty$ ones.

Let $\phi^\delta_\omega(\rho)$ be the solution that satisfies
\begin{align}\label{eq: phi delta}
  \phi^\delta_\omega(\rho_\infty -\delta) = \frac{R}{L},\qquad
  \phi^\delta_\omega(-\rho_\infty + \delta) = 0,
\end{align}
which corresponds to $\phi_\omega(\rho)$ in $R\to \infty$.
For sufficiently large $R$, we can express $\phi^\delta_\omega$ in terms of $\phi_\omega(\rho)$ and $\phi_\omega(-\rho)$, because \eqref{eq: Schrodinger} is invariant under $\rho\to -\rho$:
\begin{align}\label{eq: delta and nondelta}
  \phi^\delta_\omega(\rho) = \left(1-\frac{L^4 \omega^2}{2R^2} + \mathcal{O}(R^{-4})\right)\phi_\omega(\rho) &-\frac{L^6 B(\omega)}{R^3 A(\omega)} \left(1 + \mathcal{O}(R^{-2}) \right)\phi_\omega(\rho)\nonumber\\
  & - \frac{L^6}{R^3A(\omega)}\left(1 + \mathcal{O}(R^{-2}) \right)\phi_\omega(-\rho),
\end{align}
which satisfies the boundary conditions \eqref{eq: phi delta} up to relative errors of $\mathcal{O}(R^{-4})$.\footnote{Using \eqref{eq: delta and nondelta} for $\phi^\delta_\omega$, one finds that the field value at $\rho = \rho_\infty - \delta$ is $R/L+\mathcal{O}(R^{-3})$ and the value at the left is $0+\mathcal{O}(R^{-4})$.}

Analogously to \eqref{eq: def of N}, we define
\begin{align}\label{eq: def of N delta}
  N^\delta_n(\rho) := -\oint_{\omega_n} \frac{\d\omega}{2\pi }\phi^\delta_\omega(\rho) = \oint_{-\omega_n}\frac{\d \omega}{2\pi}\phi^\delta_\omega(\rho),
\end{align}
for which the expansion \eqref{eq: delta and nondelta} implies
\begin{align}
  N^\delta_n(\rho_\infty -\delta) = N^\delta_n(-\rho_{\infty}+\delta) = \mathcal{O}(R^{-5}).
\end{align}
Note that in \eqref{eq: def of N delta} the $\omega_n$ are the poles of $A(\omega)$ (the original eigenfrequencies of $N_n$).
Since we only need the asymptotics to the accuracy required for evaluating the action at $R\to\infty$, this definition of the bound solutions $N^\delta_n$ suffices.

\subsubsection{Solutions on SvR spacetime with $J^b \equiv 0$}\label{subsubsec: vanishing J^b}
Because the action is quadratic in $\Phi$, solutions superpose.
We therefore construct the solution with $J^b\equiv 0$ and that with $J^f\equiv 0$ separately, and add them at the end to meet \eqref{eq: soliton perturbed BC}.
Here we first find the $J^b=0$ solution.

From the above, the perturbation on $M_f$ can be written as
\begin{align}\label{eq: Feynman Phi1}
  \Phi^f(t,\rho) &= \int_{\mathrm{F}}\frac{\d\omega}{2\pi}\, e^{-i\omega t} J^f_{\omega}\phi^\delta_{\omega}(\rho) + \sum_{n = 0}^\infty \left(C_n^+ e^{-i \omega_n t} + C_n^- e^{i \omega_n t} \right)N^\delta_n(\rho),
\end{align}
where we have already used the boundary conditions \eqref{eq: rho BC} with \eqref{eq: phi delta}, and have defined
\begin{align}\label{eq: J Fourier}
  J^{f,b}_\omega &:= \int \d t\; J^{f,b}(t) e^{i\omega t}.
\end{align}
The contour $\mathrm{F}$ is the Feynman contour, shifting the poles as $\pm \omega_n \to \pm (\omega_n -i\epsilon)$ with $\epsilon\to+0$.
It is straightforward to deform to the retarded contour $\mathrm{R}$ ($\omega_n\to \omega_n-i\epsilon$) or the advanced contour $\mathrm{A}$ ($\omega_n\to \omega_n+i\epsilon$):
\begin{align}\label{eq: retarded Phi1}
  \Phi^f(t,\rho) &= \int_{\mathrm{R}}\frac{\d\omega}{2\pi}\, e^{-i\omega t}J^f_{\omega}\phi^\delta_{\omega}(\rho) + \sum_{n = 0}^\infty \left(C_n^+ e^{-i \omega_n t} + (C_n^- + J_{-\omega_n}^1) e^{i \omega_n t} \right)N^\delta_n(\rho)
  \\
  &= \int_{\mathrm{A}}\frac{\d\omega}{2\pi}\, e^{-i\omega t} J^f_{\omega}\phi^\delta_{\omega}(\rho) + \sum_{n = 0}^\infty \left((C_n^+ + J^f_{\omega_n})e^{-i \omega_n t} + C_n^- e^{i \omega_n t} \right)N^\delta_n(\rho).
  \label{eq: advanced Phi1}
\end{align}
This means that any difference among contours can be absorbed into the normalizable modes, and we will use whichever form is convenient.
To determine $C_n^\pm$, we connect the fields across $M_f\to M_b\to M_e$ using the junction conditions \eqref{eq: junction conditions} and impose periodicity on $\Sigma_{ef}$.

First we impose \eqref{eq: junction conditions} on $\Sigma_{fb}$.
From the quadratic action \eqref{eq: soliton action expansion}, continuity of the conjugate momentum of $\Phi^f$ is equivalent to continuity of $\partial_t\Phi$.
Notice that we have
\begin{align}\label{eq: Phi Sigmafb in soliton}
  \int_{\mathrm{A}}\frac{\d\omega}{2\pi}\, e^{-i\omega t_c} J^f_{\omega}\phi^\delta_{\omega}(\rho) = 0,
  \qquad
 -i  \int_{\mathrm{A}}\frac{\d\omega}{2\pi}\, \omega e^{-i\omega t_c} J^f_{\omega}\phi^\delta_{\omega}(\rho) = 0,
\end{align}
as shown in App.~\ref{app: junction}.
This implies, in \eqref{eq: advanced Phi1}, only $\{N^\delta_n(\rho)\}$ propagate into $M_b$, and hence we obtain
\begin{align}
  \Phi^b(t,\rho) = \sum_{n = 0}^\infty \left((C_n^+ + J^f_{\omega_n})e^{-i \omega_n t} + C_n^- e^{i \omega_n t} \right)N^\delta_n(\rho).
\end{align}
Next, the junction on $\Sigma_{be}$ is satisfied simply by analytic continuation $t\to -i\tau$ applied to $\Phi^b$:
\begin{align}
  \Phi^e(\tau,\rho) = \sum_{n = 0}^\infty \left((C_n^+ + J^f_{\omega_n})e^{- \omega_n \tau + i \omega_n t_c} + C_n^- e^{ \omega_n \tau - i \omega_n t_c} \right)N^\delta_n(\rho).
\end{align}
Finally, we impose periodicity on $\Sigma_{ef}$.
There, $\Phi^f$ satisfies relations analogous to \eqref{eq: Phi Sigmafb in soliton},
\begin{align}\label{eq: Phi Sigma_ef in soliton}
  \int_{\mathrm{R}}\frac{\d\omega}{2\pi}\, e^{i\omega t_c}J^f_{\omega}\phi^\delta_{\omega} = 0,
  \qquad
  -i \int_{\mathrm{R}}\frac{\d\omega}{2\pi}\, \omega e^{i\omega t_c}J^f_{\omega}\phi^\delta_{\omega} = 0.
\end{align}
Using these, we obtain
\begin{align}
  C_n^\pm = - J^f_{\pm \omega_n} n_{-\omega_n},\qquad
  n_{\omega} := \frac{1}{1-e^{-\beta \omega}}.
\end{align}

\subsubsection{Solutions on SvR spacetime with $J^f \not \equiv 0$ and $J^b \not \equiv 0$}
Repeating the same steps for $J^f\equiv 0$ and adding the two solutions yields $\Phi^{f,b}$ that satisfy \eqref{eq: soliton perturbed BC}:
\begin{align}
  \Phi^f(t,\rho) &= \int_{\mathrm{F}}\frac{\d \omega}{2\pi}\, e^{-i\omega t} J^f_{\omega}\phi^\delta_{\omega}(\rho) 
  - \sum_{n=0}^\infty\Bigl[(J^f_{\omega_n}n_{-\omega_n} + J^b_{\omega_n}e^{-\beta \omega_n}n_{\omega_n})e^{-i\omega_n t}
  \nonumber\\
  & \hspace{160pt}
   + (J^f_{-\omega_n}n_{-\omega_n} + J^b_{-\omega_n}n_{\omega_n})e^{i\omega_n t}  \Bigr]N^\delta_n(\rho),\\
  \Phi^b(t,\rho) &= \int_{\mathrm{F}}\frac{\d \omega}{2\pi}\, e^{-i\omega t} J^b_{\omega}\phi^\delta_{\omega}(\rho)
  - \sum_{n=0}^\infty\Bigl[(J^b_{\omega_n}n_{\omega_n} + J^f_{\omega_n}e^{\beta \omega_n}n_{-\omega_n})e^{-i\omega_n t}
    \nonumber\\
  & \hspace{160pt}
   + (J^b_{-\omega_n}n_{\omega_n} + J^f_{-\omega_n}n_{-\omega_n})e^{i\omega_n t}  \Bigr]N^\delta_n(\rho).
\end{align}
The Euclidean segment $M_e$ does not affect the on-shell action, so it is omitted.

Evaluating the full action using the above solutions (see \eqref{eq: soliton action expansion}, \eqref{eq: phi expansion}, \eqref{eq: delta and nondelta}, and \eqref{eq: def of N delta}) and discarding terms vanishing as $R\to\infty$, we find
\begin{align}
  \ln Z[J] &:=iS_{\mathrm{NG}}[X_f] - iS_{\mathrm{NG}}[X_b] - S_{\mathrm{NG}}[X_e]\nonumber\\
   &=-i \frac{r_*^2}{L^2}\int_{-t_c}^{t_c}\d t\;\left(J^f(t) - J^b(t) \right) + \int_{\mathrm{F}}\frac{\d \omega}{2\pi }\left(
   \frac{iR\omega^2}{2}
   +
   \frac{3iL^2B(\omega)}{2A(\omega)}
    \right)
   \left(J^f_{-\omega}J^f_{\omega} - J^b_{-\omega}J^b_{\omega} \right)
  \nonumber\\
  &\hspace{24pt}
  +\sum_{n=0}^\infty y_n \Bigl[n_{-\omega_n} J^f_{-\omega_n} J^f_{\omega_n} - n_{\omega_n} J^b_{-\omega_n} J^b_{\omega_n}
  +e^{-\beta \omega_n}n_{\omega_n} J^f_{-\omega_n}J^b_{\omega_n} + n_{\omega_n} J^b_{-\omega_n}J^f_{\omega_n}\Bigr],
  \label{eq: soliton action just evaluated}
\end{align}
where $y_n:=-3L^2B(\omega_n)/A'(\omega_n)$ and we have dropped $J$-independent constants.
As usual, we below ignore the remaining divergent kinetic term, $R\omega^2/2$.

Using \eqref{eq: ra sources} and the retarded contour, this can be recast as
\begin{align}
	\ln Z[J] &= -i \frac{r_*^2}{L^2}\int_{-t_c}^{t_c}\d t\;J^a(t) - i \int_\mathrm{R}\frac{\d \omega}{2\pi}J^a_{-\omega}\left(-\frac{3L^2 B(\omega)}{A(\omega)} \right)J^r_{\omega}\nonumber\\
	&\hspace{60pt}- \frac{1}{2}\sum_n J^a_{-\omega_n}\left(y_n\coth\frac{\beta \omega_n}{2} \right)J^a_{\omega_n}.
  \label{eq: ra S2 in soliton}
\end{align}
From this, the retarded and symmetric Green’s functions are identified as
\begin{align}
  G_\mathrm{R}(\omega) &= -3L^2 B(\omega) \mathcal{P}\frac{1}{A(\omega)} -i\pi \sum_n y_n\left[\delta(\omega-\omega_n) - \delta(\omega+\omega_n) \right],\\
  G_{\mathrm{S}}(\omega) &=i\pi \coth \frac{\beta \omega}{2}\sum_n y_n\left[\delta(\omega-\omega_n) - \delta(\omega+\omega_n) \right],
\end{align}
where $\mathcal{P}$ denotes the principal value.
In particular, the fluctuation-dissipation relation holds:
\begin{align}\label{eq: FD theorem}
  G_{\mathrm{S}}(\omega) = -\coth\!\left(\frac{\beta\omega}{2}\right)\mathrm{Im}\,G_{\mathrm{R}}(\omega).
\end{align}

\section{Deconfinement phase: a single trailing quark}\label{sec: high temp}
In this section, we consider the steady Nambu-Goto string solution \cite{Herzog:2006gh, Gubser:2006bz} in the deconfinement phase (the AdS$_5$ black brane).
As the zeroth order of the perturbation, we first construct the trailing string solution \cite{Herzog:2006gh, Gubser:2006bz} on the black brane.
We then deform the string endpoint and evaluate the on-shell action to quadratic order in the perturbation.

\subsection{Nonequilibrium steady solution}\label{subsec: steady solution}
\begin{figure}
	\centering
	\includegraphics[height = 5cm]{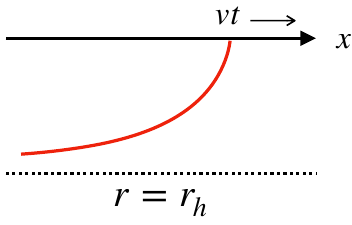}
	\caption{Configuration on a constant-$t$ slice of the solution \eqref{eq: BB xi} with $X^1 = vt - \xi(r)$.}
	\label{fig: string spiral}
\end{figure}

Following \cite{Herzog:2006gh, Gubser:2006bz}, we first construct a steady solution on the black brane in Lorentz signature.
We then use this steady solution to assemble the SvR spacetime.
This furnishes the zeroth order in the perturbative analysis in the following subsections.

The AdS$_5$ black brane solution in Lorentzian signature is
\begin{align}\label{eq: Lorentz BB}
    \d s^2 &=  \frac{r^2}{L^2}\left(-h(r)\d t^2 + (\d x^1)^2+ (\d x^2)^2+ (\d x^3)^2 \right) + \frac{L^2}{r^2h(r)}\d r^2
  \qquad
  \left(h(r) := 1 - \frac{r_h^4}{r^4} \right).
\end{align}
We put the following ansatz and solve for $\xi$ with $0<v<1$:
\begin{align}\label{eq: X^1 at 0th}
  X^1 (t,r) = v t - \xi(r),\qquad
  X^2(t,r) = X^3(t,r) = 0,\qquad
  \xi (r\to \infty) = 0.
\end{align}
The Lagrangian density of the action \eqref{eq: world sheet} then becomes
\begin{align}\label{eq: BB Lagrangian}
  \mathcal{L}_{\mathrm{NG}} = -\sqrt{1 - \frac{v^2}{h(r)} + \frac{ r^4h(r) \xi'(r)^2}{L^4}},
\end{align}
so $\xi$ is a cyclic coordinate.
Integrating the EOM once, we obtain, with an integration constant $F$,
\begin{align}\label{eq: xi prime in BB}
  \xi'(r) = -\frac{FL^4}{r^2h(r)}\sqrt{\frac{h(r)-v^2}{r^4h(r)-F^2 L^4}}.
\end{align}

We are interested in the case where $\xi(r)$ is injective; the string does not return to the boundary.
Thus the string must not have a turning point in the bulk \cite{Gubser:2006bz}.\footnote{
Imposing that the value in the square-root in \eqref{eq: BB Lagrangian} remains positive everywhere yields an equivalent condition \cite{Herzog:2006gh}.
See also \cite{Ishigaki:2023wqv} for a comprehensive study of regularity of this type.}
For this to hold, the numerator and denominator under the square root must change sign simultaneously.
From this requirement we fix $F$ as
\begin{align}\label{eq: F}
  F = \frac{r_*^2}{L^2}\sqrt{h(r_*)} = \frac{r_h^2}{L^2}\frac{v}{\sqrt{1-v^2}}
  \qquad \left(r_* := \frac{r_h}{\left(1-v^2 \right)^{1/4}} \right).
\end{align}
In terms of the temperature $T := r_h/(\pi L^2)$, the solution is given as (Fig.~\ref{fig: string spiral})
\begin{align}\label{eq: BB xi}
  \xi(r) = \frac{v}{2\pi T}\left[\tanh^{-1}\left(\frac{\pi L^2 T}{r} \right)-\tan^{-1}\left(\frac{\pi L^2 T}{r} \right) \right].
\end{align}

The induced worldsheet metric reads
\begin{align}
  \d s_{\mathrm{WS}}^2 = -(1-v^2)\frac{r^4-r_*^4}{L^2r^2}\d t^2 + \frac{L^2r^2(r^4-(1-v^2)r_h^4)}{(r^4-r_h^4)^2}\d r^2 + 2 \frac{v^2r_h^2r^2}{r^4-r_h^4}\d t\d r.
\end{align}
Introducing a new time coordinate,
\begin{align}\label{eq: hat t}
  \hat t &:= t - \frac{v^2}{1-v^2}\int_\infty^r\d r\; \frac{L^2 r_h^2 r^4}{(r^4 - r_*^4)(r^4-r_h^4)}\nonumber\\
   &= t + \frac{L^2}{2r_h}\frac{1 - (1-v^2)^{1/4}}{(1-v^2)^{1/4}} \left[\tanh^{-1}\left(\frac{r_*}{r} \right) - \tan^{-1}\left(\frac{r_*}{r} \right) \right],
\end{align}
we can diagonalize the worldsheet metric as
\begin{align}\label{eq: BB world sheet}
  \d s_{\mathrm{WS}}^2 = - (1-v^2)f(r)\d \hat t^2 + \frac{\d r^2}{f(r)}=: \gamma_{ab}\d \sigma^a \d \sigma^b
  \qquad
  \left(f(r) := \frac{r^4 - r_*^4}{L^2 r^2} \right).
\end{align}
It follows that the worldsheet is static with respect to $\hat t$ and has an effective temperature $T_* := (1-v^2)^{1/4}T$.

Below we adopt the worldsheet coordinates $\sigma = (\hat t , r)$.
Also note that as $r\to \infty$, $\hat t$ coincides with the background (and boundary) time $t$.

\subsubsection{How to define this nonequilibrium steady state and construct $M_\mathrm{SvR}$?}\label{subsubsec: NESS}
As in Sec.~\ref{sec: low temp}, we would like to obtain perturbative solutions on $M_\mathrm{SvR}$, but already at the zeroth order there is a subtlety in the construction of $M_\mathrm{SvR}$.

Let us first reconsider why the construction in Sec.~\ref{sec: low temp} worked.
There, at zeroth order, the quark-antiquark pair is at rest, and in the bulk the string takes a static configuration.
More precisely, it is along the timelike Killing vector $\partial_t$ of the AdS$_5$ soliton that the string is static.
Even if one solved the backreaction to the AdS$_5$ soliton exactly, the metric would only depend on spatial coordinates, and $t$ would remain a static time.
Therefore the quark pair and the CFT can be regarded as in mutual equilibrium, and analytic continuation to Euclidean signature can be performed without contradiction.

By contrast, the trailing string treated here is not static along the black brane time $t$.
Indeed, taking the background response seriously, the combined string-gravity system would never become stationary.
The fact that the worldsheet nevertheless admits a static time $\hat t$ is due to treating the string as a probe.
That is, the CFT remains in thermal equilibrium, thereby providing steady friction on the quark, which balances the external force acting on the quark and thus allows the nonequilibrium steady solution \eqref{eq: X^1 at 0th}.

In general, there is no guarantee that a nonequilibrium steady state (NESS) in an open-system effective theory can be described by a Euclidean path integral.
In the present case it is also difficult to do so purely from the boundary perspective.
We therefore exploit the fact that, in the probe approximation, the bulk theory reduces to the Nambu-Goto action alone, and we define this NESS holographically from the bulk string path-integral viewpoint:
\begin{align}\label{eq: NESS}
  \braket{X_L|\rho_{\mathrm{NESS}}(T_*)|X_R} = \int_{X(0,r) = X_R(r)}^{X(\beta_*,r)=X_L(r)} \mathcal{D}X e^{-S_{\mathrm{NG}}[X;0\le \hat \tau \le \beta_*]}\delta(X(\hat \tau, \infty) - x_{\mathrm{bdy}}(\hat \tau) ).
\end{align}
Here $\hat \tau = i\hat t$, $\beta_* := T_*^{-1}$, and $x_{\mathrm{bdy}}$ denotes the boundary worldline of the quark:
\begin{align}
  x_{\mathrm{bdy}}(\hat \tau) = (x_{\mathrm{bdy}}^{\hat \tau}(\hat \tau),x_{\mathrm{bdy}}^{1}(\hat \tau),x_{\mathrm{bdy}}^{2}(\hat \tau),x_{\mathrm{bdy}}^{3}(\hat \tau)) := (\hat \tau, iv \hat \tau,0,0).
\end{align}
As $r\to \infty$, $X_R$ and $X_L$ coincide with $x_{\mathrm{bdy}}(0)$ and $x_{\mathrm{bdy}}(\beta_*)$, respectively.
We take $(\hat \tau, r) \in [0,\beta_*]\times [r_*, \infty]$ as the worldsheet coordinates and gauge-fix as $X^{\hat \tau} = \hat \tau$, but we do not require $X^r = r$; the integration measure should retain the freedom regarding how far the string extends into the bulk.

Since, in the probe approximation, the gravitational degrees of freedom remain approximately in thermal equilibrium, the total bulk state can be written formally as
\begin{align}\label{eq: bulk stete}
  \rho_{\mathrm{tot}} \simeq \rho_{\mathrm{NESS}}(T_*) \otimes e^{- H_{\mathrm{grav}}/T}/Z_{\mathrm{grav}}(T)
  \qquad
  \left(Z_{\mathrm{grav}}(T) := \mathrm{Tr}\,e^{- H_{\mathrm{grav}}/T} \right)
\end{align}
with the background temperature staying at $T$.

The relation between \eqref{eq: bulk stete} and \eqref{eq: BB world sheet} becomes clear by evaluating the partition function in \eqref{eq: bulk stete}.
The gravitational partition function reduces to $Z_{\mathrm{grav}}(T)$ and is realized by the Euclidean version of \eqref{eq: Lorentz BB} as the dominant saddle in the high temperature phase.
On the other hand, from \eqref{eq: NESS}, the string partition function reads
\begin{align}\label{eq: NESS partition}
  Z_{\mathrm{NESS}}(T_*) :=  \oint \mathcal{D}X e^{-S_{\mathrm{NG}}[X;0\le \hat \tau \le \beta_*]}\delta(X(\hat \tau, \infty) - x_{\mathrm{bdy}}(\hat \tau) ).
\end{align}
Since the direction along which $x_{\mathrm{bdy}}$ looks static is $\partial_{\hat \tau} + iv \partial_{x^1}$, the periodic boundary condition must be imposed along this direction.
Accordingly, $X^1$ is twisted, $X^1(\beta_*,r) = X^1(0,r) + iv \beta_*$, as we do in Wick rotation of spinning black holes.
The classical solution obtained from \eqref{eq: X^1 at 0th} with $\hat t \to -i\hat \tau$ obviously satisfies this boundary condition\footnote{
In doing $\hat t \to -i\hat \tau$, we regard the $t$ in \eqref{eq: X^1 at 0th} as the function of $\hat t$ and $r$ given via \eqref{eq: hat t}.} and, by \eqref{eq: BB world sheet}, has no conical singularity with imaginary-time period $\beta_*$, so it is a saddle of \eqref{eq: NESS partition}.

Therefore, in the probe limit, one can construct the zeroth-order SvR spacetime using only the worldsheet, and \eqref{eq: BB world sheet} dominates in the classical approximation.
Here, the time $\hat t$ runs the contour of Fig.~\ref{fig: SK contour} with $\beta\to \beta_*$, and accordingly, \eqref{eq: BB world sheet} constructs the ``SvR worldsheet" as in the right panel of Fig.~\ref{fig: SvR spacetime}.
In this way the junction conditions \eqref{eq: junction conditions} are naturally satisfied at zeroth perturbative order.
As in Sec.~\ref{sec: low temp}, the perturbative solutions below are treated as a scalar field theory living on this SvR worldsheet.\footnote{
In \cite{Bu:2021jlp}, the analytic-continuation prescription of \cite{Glorioso:2018mmw} is applied to the effective worldsheet horizon.
Although its validity is not discussed there, doing so automatically guarantees that the generating functional obeys the KMS condition.
This implicitly assumes that an approximate thermal steady state focused on the quark can be defined.}

\subsection{Quadratic effective action up to second order in perturbations}\label{subsec: quadratic action high temp}
With the preparations complete, we now switch on perturbations as
\begin{align}\label{eq: perturbation Phi}
  &\delta X^1_{f,b,e} = \Phi^{f,b,e}(\hat t,r)\\
   &\mbox{with}\quad
  \Phi^f \sim  J^f,\quad
  \Phi^b \sim J^b,\quad
  \Phi^e \sim 0\quad
  (\mbox{as }r\to \infty),
  \label{eq: Phi BC}
\end{align}
and evaluate the action to quadratic order in $J$.
On $M_f$, the perturbative expansion of the action is given as
\begin{align}\label{eq: BB NG expansion}
  S_{\mathrm{NG}}[X_f] = (\mathrm{const})  -F\int_{-t_c}^{t_c}\d t\; J^f(t) - \frac{1}{2\sqrt{1-v^2}} \int_{r_*}^{\infty} \d r\int_{-t_c}^{t_c} \d \hat t\, g_{11} \gamma^{ab}\partial_{a} \Phi \partial_{b}\Phi
 \end{align}
where $\gamma_{ab}$ is given by \eqref{eq: BB world sheet}, and $g_{11} = r^2/L^2$ is the $x^1x^1$-component of the background metric \eqref{eq: Lorentz BB}.

\subsubsection{EOM and general solution in Lorentz signature}
With the Fourier representation \eqref{eq: Fourier tr}, the EOM for $\Phi^f$ reads
\begin{align}\label{eq: BB EOM}
	\frac{1}{1-v^2} \frac{r^4}{L^4\hat f(r)} \omega^2 \Phi^f_\omega(r) + (\hat f(r)\Phi^f_\omega\.'(r))' = 0\qquad
	\left(\hat f(r) := \frac{r^4 - r_*^4}{L^4} \right)
\end{align}
where the prime denotes $\partial_r$.

Two independent solutions of \eqref{eq: BB EOM} are classified by their behavior near the effective horizon $r\to  r_*$:
\begin{align}\label{eq: phi is ingoing}
  \phi_{\pm \omega}(r) \sim \mathrm{const}\cdot \left(r-r_* \right)^{\mp i a \omega}
  \qquad
  \left(a := \frac{L^2}{4r_*\sqrt{1-v^2}} \right).
\end{align}
If we denote one independent solution by $\phi_{\omega}$, the symmetry $\omega \leftrightarrow - \omega$ of \eqref{eq: BB EOM} implies that the other is $\phi_{-\omega}$.
On the other hand, we can also classify the solutions according to the behavior at $r\to \infty$, and we find $r^{0}$ and $r^{-3}$.
Therefore $\phi_{\omega}$ must contain a component of the $O(r^0)$ mode; otherwise any linear combination of $\phi_{\pm \omega}$ would produce only the $O(r^{-3})$ mode.
We thus can normalize $\phi_{\omega}$ as
\begin{align}\label{eq: phi normalization}
  \phi_{\omega} \overset{r\to \infty}{\to} 1,
\end{align}
and define the $O(r^{-3})$ mode by
\begin{align}\label{eq: BB normalizable mode}
  N_\omega(r) := \phi_{\omega}(r)-\phi_{-\omega}(r).
\end{align}
The higher order correction to\eqref{eq: phi normalization} is calculated as
\begin{align}\label{eq: BB asymptotics}
  \phi_{\omega} \overset{r\to \infty}{\sim} 1 + \frac{L^4 \omega^2}{2(1-v^2)}r^{-2} + A(\omega) r^{-3} + \cdots
  \qquad
  (\mbox{as $r\to \infty$}),
\end{align}
where $A(\omega)$ is determined from the boundary condition near the horizon, \eqref{eq: phi is ingoing}.

Finally, let $\phi^R_\omega$ be the solution that is exactly $1$ at the cutoff surface $r = R$.
Then we find
\begin{align}
  \phi^R_\omega(r) := \frac{\phi_{\omega}(r)}{\phi_{\omega}(R)} \overset{r\to  R}{\sim} 1 + \frac{L^4\omega^2}{2(1-v^2)}(r^{-2} - R^{-2} ) + A(\omega) (r^{-3} - R^{-3}) + \mathcal{O}(R^{-4}).
\end{align}
In the analysis below, a delicate order estimation is unnecessary, and—as is common in the literature—we may work directly with $\phi_\omega$.
Readers concerned about this point may replace $\phi_\omega$ by $\phi_\omega^R$ in what follows and verify that nothing essential changes.

\subsubsection{Solution on the SvR spacetime with $J^b \equiv 0$}
As before, we first set $J^b \equiv 0$.
A general solution for $\Phi_{\omega}$ is a linear combination of $\phi_{\pm \omega}$.
Hence, on $M_f$, a general solution satisfying the boundary condition $\Phi^f \to J^{f}$ is
\begin{align}\label{eq: Phi M1 in BB}
  \Phi^f = \int\frac{\d\omega}{2\pi}\, e^{-i\omega \hat t}\left(C_{\omega} \phi_{\omega}(r) + \left(J^f_{\omega} - C_{\omega} \right)\phi_{-\omega}(r) \right),
\end{align}
where $C_{\omega}$ is an arbitrary constant and $J^f_\omega$ is defined in \eqref{eq: J Fourier}.
The fulfillment of the boundary condition $\Phi \to J^f$ is immediate from the normalization \eqref{eq: phi normalization}.

In the black hole background, poles are absent on the real axis, so one does not need the $i\epsilon$-prescription.
Since relations analogous to \eqref{eq: Phi Sigmafb in soliton} and \eqref{eq: Phi Sigma_ef in soliton} hold\footnote{
Here we use the outgoing mode $\phi_{-\omega}$ in place of the advanced contour and the ingoing mode $\phi_{\omega}$ in place of the retarded contour.
For the derivation, see App.~\ref{app: junction}.}, precisely the same steps as in Sec.~\ref{subsubsec: vanishing J^b} give
\begin{align}\label{eq: Phi Mb in BB}
  \Phi^b(\hat t,r) &= \int\frac{\d\omega}{2\pi}\, e^{-i \omega \hat t}\, C_{\omega} N_{\omega}(r)\\
  \Phi^e(\hat \tau,r) &= \int\frac{\d\omega}{2\pi}\, e^{i \omega t_c - \omega \hat \tau}\, C_{\omega} N_{\omega}(r)\\
  C_\omega &= J^f_\omega n_\omega
  \qquad
  \left(n_\omega := \frac{1}{1 - e^{-\beta_* \omega}} \right).
\end{align}

\subsubsection{Solution on the SvR spacetime with $J^f\not \equiv 0$ and $J^b\not \equiv 0$}
Tracing the same steps with $J^f\equiv 0$ gives the companion solution.
Adding the two yields a solution that satisfies \eqref{eq: Phi BC}:
\begin{align}
  \Phi^f(\hat t,r) &= \int\frac{\d\omega}{2\pi}\, e^{-i\omega \hat t}n_\omega \left[(J_\omega^f - e^{- \omega \beta_*} J^b_\omega)\phi_\omega(r)
  -e^{-\omega \beta_*}(J^f_\omega - J^b_\omega)\phi_{-\omega}(r)
   \right],
  \\
  \Phi^b(\hat t,r) &= \int\frac{\d\omega}{2\pi}\, e^{-i\omega \hat t}n_\omega \left[
  (J^f_\omega - e^{-\omega \beta_*}J^b_\omega)\phi_\omega(r) - (J^f_\omega - J^b_\omega)\phi_{-\omega}(r)
   \right].
\end{align}
Again, we omit $\Phi^e$, which does not contribute to the on-shell action.

We now evaluate the action.
From the asymptotic expansion \eqref{eq: BB asymptotics}, we have
\begin{align}\label{eq: H in BTZ}
  -\frac{1}{\sqrt{1-v^2}}g_{11}(r) \gamma^{rr} \phi_{\omega}'(r) \overset{r\to \infty}{\sim}  \frac{R}{L^2(1-v^2)^{3/2}}L^2 \omega^2 + \frac{3A(\omega)}{L^4\sqrt{1-v^2}} =: H_\omega
\end{align}
where $r=R$ is the boundary cutoff surface and terms vanishing as $R\to \infty$ are dropped.\footnote{
$M_v :=  R/L^2(1-v^2)^{3/2}$ can be interpreted as the mass of a heavy boundary particle and appears in the Langevin equation \cite{Son:2009vu}.}
Using these, the full action evaluates to\footnote{
No surface term arises from the side $r\to r_*$: $\Phi^{f,b,e}(r)$ and $(r-r_*)\Phi^{f,b,e}\.'(r)$ vanish by the Riemann-Lebesgue lemma, because $\phi_{\pm\omega}$ oscillate rapidly near $r\to r_*$.}
\begin{align}
  \ln Z[J] &:= i S_{\mathrm{NG}}[X_f] -  i S_{\mathrm{NG}}[X_b] - S_{\mathrm{NG}}[X_e]\nonumber\\
  	&= - iF\int_{-t_c}^{t_c}\d \hat t\; \left(J^f(t) - J^b(t) \right)\nonumber\\
  	&\hspace{36pt}
  	+ i\int \frac{\d \omega}{2\pi}\left[J^f_{-\omega} H_\omega n_\omega J^f_\omega - J^b_{-\omega} H_{-\omega} n_\omega J^b_\omega - J^b_{-\omega}(H_\omega -  H_{-\omega})n_{\omega}J^f_\omega \right]
\end{align}
where we have discarded $J$-independent constants.
Introducing \eqref{eq: ra sources}, we rewrite this as
\begin{align}
  \ln Z[J] = - iF\int_{-t_c}^{t_c}\d \hat t\; J^a(t) + \int\frac{\d\omega}{2\pi}\left[-i J^a_{-\omega}(-H_{\omega})J^r_\omega - \frac{1}{2}J^a_{-\omega}\left(\frac{H_\omega - H_{-\omega}}{2i}\coth\frac{\omega \beta_*}{2} \right)J^a_{\omega} \right].
  \label{eq: ra S2 in BB}
\end{align}

From the above action, after removing the divergent mass term, the retarded and symmetric Green’s functions are read as
\begin{align}
  G_{\mathrm{R}}(\omega) &= -\frac{3A(\omega)}{L^4\sqrt{1-v^2}}\\
  G_{\mathrm{S}}(\omega) &= \frac{3(A(\omega) - A(-\omega))}{2iL^4\sqrt{1-v^2}}\coth\frac{\omega \beta_*}{2}.
\end{align}
They obviously satisfy the fluctuation-dissipation relation \eqref{eq: FD theorem}.

\section{Summary and discussion}\label{sec: discussion}
In this paper we perturbatively constructed string solutions on the SvR spacetime around steady configurations in both the confinement phase (the AdS$_5$ soliton) and the deconfinement phase (the AdS$_5$ black brane), and evaluated the on-shell action up to quadratic order in perturbations.
In particular, the computation in the confinement phase was made possible precisely because the SvR prescription provides a holographic SK framework that does not assume the presence of a horizon.
For the deconfinement phase, \cite{Bu:2021jlp} has already carried out the cubic order analysis using \cite{Glorioso:2018mmw}, while in this paper revisited the problem to a quadratic order with the SvR prescription.

Analytic expressions for $A(\omega)$, $B(\omega)$, and the $\omega_n$ that appear in the Green functions are difficult to obtain, and as stated in the introduction, the analysis of their properties is left for future work.
These data reduce to the eigenvalue problem of the Schr\"odinger equation \eqref{eq: Schrodinger} and to the asymptotic analysis of the solution $\phi_\omega$.
For the eigenvalue problem in AdS, \cite{Ishii:2015wua} performed numerical analyses for various perturbations.
Moreover, \cite{Bu:2021jlp} computed the low-energy expansion of the Green function and numerical plots in the deconfinement phase, which will also be a useful reference.

As mentioned in the introduction, for the Brownian motion of a quark in the deconfinement phase, its Langevin dynamics has been extensively studied.
We can generalize this classical concept to quantum one, by regarding \eqref{eq: ra S2 in BB} as the influence functional for a quantum Brownian particle.
By doing so, a Lindblad equation for one dimensional holographic Brownian particle is derived in \cite{TakedaHQBM}.\footnote{
A treatment of the Lindblad master equation when the CFT is the target open system has been proposed in \cite{Ishii:2025qpy}.}
If the transverse sectors ($x^{2,3}$-directions) is also taken into account, we may be able to directly compare the Lindbladian with that of the quark-gluon plasma in QCD.
Such a master equation is proposed, for example, in \cite{Akamatsu:2014qsa}.

It is also important to understand the interrelations among the various holographic SK prescriptions proposed so far.
The SvR prescription is powerful when the closed-time-path integral can be written from the boundary perspective; if so, one can deduce what must be computed in the bulk by directly using the AdS/CFT dictionary.
However, this requires that the initial equilibrium state be representable by a Euclidean path integral, and in Sec.~\ref{subsubsec: NESS} we indeed encountered a subtlety along these lines.
On the other hand, the method of \cite{Glorioso:2018mmw} is more flexible so long as a horizon is present.
For example, when the theory varies slowly in space and time, one may approximately regard each spacetime point as locally equilibrated and define a local temperature.
In such situations, following \cite{Glorioso:2018mmw}, one may consider a slowly varying black hole spacetime to obtain a local KMS condition.
It would be desirable in the future to justify \cite{Glorioso:2018mmw} from a top-down viewpoint, or to investigate its equivalence to the SvR prescription.

\subsection*{Acknowledgment}
The authors acknowledge S. Sugimoto for discussions and comments.
The work of S.\,N.~is supported in part by JSPS KAKENHI Grant No.~JP25K07174, and the Chuo University Personal Research Grant.
D.T.\ was supported by Grant-in-Aid for JSPS Fellows No.\ 22KJ1944 and is also by RIKEN Special Postdoctoral Researchers Program.

\appendix
\section{Perturbative expansion formula for the Nambu-Goto action}\label{app: NG expansion}
\begin{figure}
	\centering
	\includegraphics{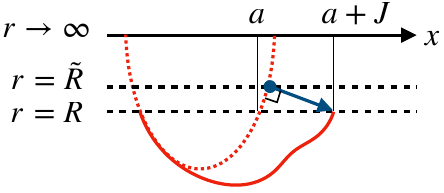}
	\caption{Mapping from the static solution to the perturbed solution.
	Since the perturbation is introduced along the normal, we extend the worldsheet coordinate $r$ to $r \le \tilde R$ so that the boundary condition at $r=R$ is satisfied.}
	\label{fig: perturbed string}
\end{figure}

In this appendix, we detail the origin of the boundary condition \eqref{eq: rho BC} and the expansion \eqref{eq: soliton action expansion}.
We focus on $M_f$, and the other segments are treated in exactly the same way.
We also focus on the right half of the string, where the perturbation is applied to the endpoint.
We use $(t,r)$ rather than $(t,\rho)$ as the worldsheet coordinates.

We begin by reconsidering the normal perturbation \eqref{eq: normal perturbation} of the static solution.
Since the normal vector $n^\mu$ appearing in \eqref{eq: normal perturbation} is a vector defined at each point on the static string, introducing the deformation via \eqref{eq: normal perturbation} amounts to fixing one gauge; equivalently, to keep using $n^\mu$ away from the string, one is forced to adopt the coordinate representation of $n^\mu$.
For a tidy expansion of the action, however, it is preferable to define the perturbation in a coordinate-covariant manner.

To this end, we define a (non-perturbative) deformation of the embedding to using the exponential map:\footnote{For a vector $V$ in the tangent space at $p$, $x(s):=\exp_p(V)$ denotes the point at affine parameter $1$ along the affine geodesic shot from $p$ with the initial velocity set $\dot x(0) = V$.}
\begin{align}\label{eq: exact deformation}
  X_\Phi(t, r) := \exp_{p}\!\left(\Phi\, n^\mu \right).
\end{align}
Here $p$ is the spacetime point on the static solution specified by the worldsheet coordinates $(t,r)$.
We only need the expansion up to $\mathcal{O}(\Phi^2)$; by definition and the geodesic equation,
\begin{align}\label{eq: exact deformation expanded}
  X_\Phi^\mu(t,r) = X_0^\mu +  \Phi\, n^\mu - \frac{1}{2}\Gamma^\mu_{\nu\rho} n^\nu n^\rho \Phi^2 + \mathcal{O}(\Phi^3),
\end{align}
where $X_0^\mu := (t,r,a-\xi(r),0,0)$ is the coordinate of $p$, and $\Gamma^\mu_{\nu\rho}$ is the Christoffel symbol of the AdS$_5$ soliton metric.
The function $\xi$ is the solution obtained in Sec.~\ref{subsec: both endpoints}.
Note that $X_\Phi^t = t$ and $X_\Phi^2=X_\Phi^3=0$ hold exactly.

We are now ready to discuss the boundary condition.
With a cutoff surface introduced at $r=R$, the right-endpoint boundary condition in \eqref{eq: soliton perturbed BC} becomes\footnote{We have shifted $\xi(r)\to \xi(r)-\xi(R)$ so that $\xi$ vanishes at $r=R$, not at $r\to \infty$.
Since $\xi$ was a “cyclic coordinate,” the shifted $\xi$ still solves the equations of motion.}
\begin{align}\label{eq: exact BC}
  X_\Phi^r (t,\tilde R(t);s) = R,\qquad
  X_\Phi^1(t,\tilde R(t);s) = a + J(t).
\end{align}
It is crucial here that we impose the boundary condition at $r=\tilde R(t)$ for some function $\tilde R(t)$.
As illustrated in Fig.~\ref{fig: perturbed string}, for $\Phi\neq 0$ we have $X^r(t,R;s)\neq R$, so the boundary of the worldsheet coordinate $r$ must be adjusted.
Eq.\ \eqref{eq: exact BC} determines both the boundary value of $\Phi$ and the function $\tilde R(t)$.

Since we are interested in a perturbative analysis in $J^f$, we treat $\Phi=\mathcal{O}(J)$ and solve \eqref{eq: exact BC} perturbatively.
Keeping only the terms that survive as $R\to\infty$, and using \eqref{eq: xi prime in soliton} and related relations, we obtain
\begin{align}
  \Phi(t,R) &= \frac{R}{L}\Bigl(J(t) + \mathcal{O}(R^{-4})\Bigr) + \frac{r_*^2}{L^3}J(t)^2 + \mathcal{O}\!\left(J(t)^3\right),
  \label{eq: Phi pre BC}\\
  \tilde R(t) &= R + \frac{r_*^2}{L^2} J(t) - \frac{R^3}{2L^4}J(t)^2 + \mathcal{O}\!\left(J(t)^3\right).
\end{align}
Among the omitted powers of $R$ in \eqref{eq: Phi pre BC}, only the $J \times \mathcal{O}(R^{-4})$ piece has to be carefully followed later, which is why we displayed it explicitly.

With these in hand, the expansion \eqref{eq: soliton action expansion} follows as we see from now.
Evaluating the Nambu-Goto action on the deformed embedding gives the exact expression
\begin{align}
  S_{\mathrm{NG}}[X_{\Phi}] = -\int_{-t_c}^{t_c}\d t \int_{r_*}^{\tilde R(t)}\d r\,\sqrt{-\gamma_{\Phi}},
\end{align}
where $(\gamma_{\Phi})_{ab}$ is the induced metric on the deformed worldsheet $x=X_\Phi$.
We denote by $\gamma_{ab}$ the induced metric before deformation.
Referring to \cite{Capovilla:1994bs, Kiosses:2014tua}, the Lagrangian density expands as
\begin{align}
  \sqrt{-\gamma_{\Phi}} = \sqrt{-\gamma} +  \frac{1}{2}\sqrt{-\gamma}\Bigl(\gamma^{ab}\partial_a \Phi \partial_b \Phi + m^2(r)\Phi^2 \Bigr) + \mathcal{O}(J^3),
\end{align}
with $m^2(r)$ given in \eqref{eq: mass of r}.
Therefore, the action itself expands as
\begin{align}\label{eq: pre soliton action expansion}
  S_{\mathrm{NG}}[X_{\Phi}] =\; &(\mathrm{const}) \;-\; \bigl(\tilde R(t)-R\bigr)\int_{-t_c}^{t_c}\d t\,\sqrt{-\gamma}\big|_{r=R} - \frac{\bigl(\tilde R(t)-R\bigr)^2}{2}\int_{-t_c}^{t_c}\d t\,\partial_r \sqrt{-\gamma}\big|_{r=R} \nonumber\\
  &- \frac{1}{2}\int_{-t_c}^{t_c}\d t \int_{r_*}^{R}\d r\,\sqrt{-\gamma}\,\Bigl(\gamma^{ab}\partial_a \Phi \partial_b \Phi + m^2(r)\Phi^2 \Bigr) + \mathcal{O}(J^3).
\end{align}
Keeping terms to $\mathcal{O}(j^2)$, which do not vanish as $R\to \infty$, we find that \eqref{eq: pre soliton action expansion} agrees with \eqref{eq: soliton action expansion}.

Finally, let us revisit \eqref{eq: Phi pre BC}.
The $\mathcal{O}(J^2)$ part of \eqref{eq: Phi pre BC} can be dropped, because it contributes only at $\mathcal{O}(J^3)$ to the action \eqref{eq: pre soliton action expansion}.
The $\mathcal{O}(J)\times \mathcal{O}(R^{-4})$ term can also be ignored.
If one insisted on keeping it, one could incorporate an $\mathcal{O}(R^{-4})$ correction on the right-hand side of the first condition in \eqref{eq: phi delta}, and then could still use the perturbative solution \eqref{eq: Feynman Phi1} as is.
However, we already know from \eqref{eq: delta and nondelta} that neglecting $\mathcal{O}(R^{-4})$ does not affect the action evaluation \eqref{eq: soliton action just evaluated}.
Thus it is safe to use the boundary condition \eqref{eq: rho BC}.

\section{Details of the junction conditions}\label{app: junction}
In this appendix we explain the relations \eqref{eq: Phi Sigmafb in soliton} and \eqref{eq: Phi Sigma_ef in soliton}, which were deferred in Secs.~\ref{sec: low temp} and \ref{sec: high temp}.
We focus on \eqref{eq: Phi Sigmafb in soliton}.
Extending the solution to $t\ge t_c$, \eqref{eq: Phi Sigmafb in soliton} can be recast as
\begin{align}\label{eq: advanced vanishes}
	\Phi_{\mathrm{A}}(t,\rho)
  :=
  \int_{\mathrm{A}} \frac{\d\omega}{2\pi}\, e^{-i \omega t} J^f_{\omega} \phi_{\omega}(\rho) \equiv 0
  \qquad
  \left(t\ge t_c \right),
\end{align}
so it suffices to show this (with the support of $J^f(t)$ kept on $[-t_c,t_c]$).
In the deconfinement phase, we omit the advanced contour $\mathrm{A}$, instead use the outgoing mode $\phi_{-\omega}$, and work with the radial coordinate $r$.

Taking into account that the support of $J^f(t)$ lies in $[-t_c,t_c]$, the left-hand side can be rearranged as
\begin{align}\label{eq: illegal transformation}
   \int_{\mathrm{A}} \frac{\d\omega}{2\pi}\, e^{-i \omega t} J^f_{\omega} \phi_{\omega}(r)
   =
   \int_{-t_c}^{t_c} \d t'\, J^f(t') 
    \int_{\mathrm{A}} \frac{\d\omega}{2\pi}\, e^{-i \omega (t-t')} \phi_{\omega}(r).
\end{align}
Since the poles are avoided by the advanced contour $\mathrm{A}$\footnote{In a black-hole spacetime, we assume no quasi-normal mode poles in the lower half plane. Otherwise they would represent unstable modes that grow in time.} and $t\ge t_c>t'$, we can deform the $\omega$ integral to the infinite semicircle in the lower half-plane, where it vanishes.

\section{SK action for scalar fields}\label{app: scalar}
There is a vast literature reporting scalar field computations based on the SvR prescription.
For future reference, we here solve the free scalar field on $M_{\mathrm{SvR}}$ made of a generic static and isometric black brane spacetime, and show the on-shell SK action.

Suppose that the background spacetime is a generic static and isometric black brane spacetime:
\begin{align}\label{eq: generic VV}
    \d s^2 &=  \frac{r^2}{L^2}\left(-h(r)e^{\alpha(r)}\d t^2 + \delta_{ij} \d x^i \d x^j\right) + \frac{L^2}{r^2h(r)}\d r^2.
\end{align}
Here, $h$ behaves as
\begin{align}
    h(r) \overset{r\to r_h}{\sim} b (r-r_h)\qquad (a>0),
\end{align}
and $\alpha$ is regular for $r\ge r_h$.

The Klein-Gordon equation in the Fourier space reads
\begin{align}\label{eq: scalar EOM}
    \frac{e^{-\alpha}}{h}\omega^2\Phi_{\omega,k} - k^2\Phi_{\omega,k} + \frac{L^3}{r^3}\left(\frac{r^5}{L^5}h \Phi_{\omega,k}' \right)' - m^2\Phi_{\omega,k} = 0,
\end{align}
where $k = (k^1,\cdots,k^{d-1})$ and ${}'$ denotes $r$-derivative.
The solutions can be classified by two independent solutions $\phi_{\pm \omega,k}$ that behaves as
\begin{align}
    \phi_{\omega,k} \overset{r\to r_h}{\sim} \mathrm{const}\cdot (r-r_h)^{-ia\omega}
    \qquad
    \left(a := \frac{Le^{-\alpha(r_h)/2}}{br_h}\; \right).
\end{align}
From the same argument as Sec.\ \ref{sec: high temp}, $\phi_{\omega, k}$ must be non-normalizable, and thus can be expanded as
\begin{align}
    \phi_{\omega, k} \overset{r \to \infty}{\sim} r^{\Delta-d}(1 + \mathcal{O}(r^{-2})) + O_{\omega,k} r^{-\Delta}(1 + \mathcal{O}(r^{-2}))
    \quad
    \mbox{with}
    \quad
    \Delta := \frac{d}{2} + \sqrt{\frac{d^2}{4}+m^2L^2},
\end{align}
with some coefficient $O_{\omega, k}$.
Note that we have fixed the normalization of $\phi_{\omega,k}$ through this expansion.
The normalizable mode can be made as $N_{\omega, k} = \phi_{\omega,k} - \phi_{-\omega,k}$.

The construction of the solution on $M_\mathrm{SvR}$ (Fig.\ \ref{fig: SvR spacetime} right) with \eqref{eq: bulk BC} is quite parallel to Sec.\ \ref{sec: high temp}.
The result is
\begin{align}
  \Phi_f &= \int\frac{\d\omega \d k}{(2\pi)^d}e^{-i\omega t + i k x}\; \left[J^f_{\omega, k}\phi_{\omega,k}+(J^f_{\omega, k}-J^b_{\omega, k})e^{-\beta\omega}n_{\omega}N_{\omega,k} \right],\\
  \Phi_b &= \int\frac{\d\omega \d k}{(2\pi)^d}e^{-i\omega t + i k x}\; \left[J^b_{\omega, k}\phi_{\omega,k}+(J^f_{\omega, k}-J^b_{\omega, k})n_{\omega}N_{\omega,k} \right],\\
  \Phi_e &= \int\frac{\d\omega \d k}{(2\pi)^d}e^{i\omega t_c -\omega \tau + i k x}\;(J^f_{\omega, k}-J^b_{\omega, k})n_{\omega}N_{\omega,k},
\end{align}
with $n_{\omega} := (1-e^{-\beta \omega})^{-1}$.
With the help of App.\ \ref{app: junction}, we see that the junction conditions are also satisfied.

By those solution, we evaluate the total action $S = S_f - S_b +i S_e$.
To do so, we have to introduce the counterterms \cite{deHaro:2000vlm} to each segment.
For $d/2<\Delta < 1 + d/2$, the following counterterm suffices:
\begin{align}
    S_{\mathrm{ct}} = \frac{\Delta - d}{2L} \int_{r=R}\d t\d^{d-1}x \sqrt{|h_R|}\; \Phi^2.
\end{align}
Here, $r = R$ is the cutoff surface and $h_{R}$ is the determinant of the induced metric there.
For such $\Delta$, we obtain, by taking $R\to \infty$,
\begin{align}
    S = \int\frac{\d\omega \d k}{(2\pi)^d}\left[
    \left(\frac{2\Delta-d}{L^{d+1}}O_{\omega,k}\right) J^r_{-\omega,-k}J^a_{\omega,k}
    - i\left(\frac{2\Delta-d}{2L^{d+1}}\coth\frac{\beta \omega}{2}\;\mathrm{Im}\;O_{\omega, k}\right)|J^a_{\omega,k}|^2
    \right].
\end{align}
The fluctuation-dissipation relation is obvious from the action.

\bibliographystyle{jhep}
\bibliography{ref}

\end{document}